# On the Analytical Tractability of Hexagonal Network Model with Random User Location

Ridha Nasri and Aymen Jaziri

*Abstract*—Explicit derivation of interferences in hexagonal wireless networks has been widely considered intractable and requires extensive computations with system level simulations. In this paper, we fundamentally tackle this problem and explicitly evaluate the downlink Interference-to-Signal Ratio (ISR) for any mobile location $m$ in a hexagonal wireless network, whether composed of omni-directional or tri-sectorized sites. The explicit formula of ISR is a very convergent series on $m$ and involves the use of Gauss hypergeometric and Hurwitz Riemann zeta functions. Besides, we establish simple identities that well approximate this convergent series and turn out quite useful compared to other approximations in literature. The derived expression of ISR is easily extended to any frequency reuse pattern. Moreover, it is also exploited in the derivation of an explicit form of SINR distribution for any arbitrary distribution of mobile user locations, reflecting the spatial traffic density in the network. Knowing explicitly about interferences and SINR distribution is very useful information in capacity and coverage planning of wireless cellular networks and particularly for macro-cells' layer that forms almost a regular point pattern.

## I. Introduction

### A. Background and Motivation

INTERFERENCE in wireless networks is the prime concern of telecommunication actors which continuously attempt to minimize it in all the stages of the technology conception and deployment (from the standardization to the process of network design, planning and exploitation). Interferences are often attributed to co-channel interferences which are the result of the scarce spectrum reuse in different cells of the network. In the network design phase, planning engineers try to understand the behavior of the system in terms of radio capacity and relate it to interferences. Often, they need to have a quick answer on its estimation without having recourse to a simulator. This answer should come up with a closed form expression of interferences, required to be directly used in some tools essential for capacity and coverage planning, such as link budget tool.

In literature, co-channel interferences are often represented by the Interference-to-Signal Ratio (ISR) metric [1], named also interference factor in 3G studies [2], [3]. It is defined as the normalized intercell interferences against the received useful signal (received power from the serving cell). ISR metric constitutes the chestnut of the performance analysis, in terms of SIR (Signal-to-Interference-Ratio) and SINR (Signal-to-Interference-plus-Noise-Ratio). Its evaluation in wireless networks has been widely addressed but it is still a fresh topic interesting both academic and industrial communities [1]–[11]. Interferences depend on the geometric placement of radio sites in the network. The two most frequently considered geometry models are hexagonal, named also equilateral triangular lattice, and random networks. The first assumes that sites are placed on a regular infinite hexagonal grid whereas the second supposes that sites are randomly distributed according to a spatial point process, e.g., Poisson Point Process [6]. Although random models have recently drawn more attention of the research community, hexagonal model is always used as the basis for coverage dimensioning (estimation of the number of sites required to cover an area with a given quality of service) and by operational staff for conducting strategic studies of a typical regular network assuming a complete sites' deployment. Despite its frequent use in network geometry modeling, hexagonal model has been widely considered intractable and the evaluation of interferences requires extensive numerical computations with simulations [12], [13]. This intractability was the key driver in the migration to random network models. Besides, system performances are strongly linked not only to the network geometry but also to user location distributions. With random models, it is hard to analytically evaluate the co-channel interference in each user location [1], only its distribution and eventually its average are determined. Moreover, random network models are often taken stationary: distributions of ISR and SINR are independent from the user location.

In this work, we confine ourselves to mathematically tackle the tractability issue of the hexagonal model considering different user location distributions. Besides, this paper finds its motivation from the limit of random network models in dealing with non-uniform user location distribution that has recently come to prominence [14].

### B. Related works

Several works have been made to give approximations of interferences in regular hexagonal networks based on observed simulation results [2]–[4], [11]. To the best of our knowledge, no closed exact formula, mathematically and rigorously proved, was found.

By way of examples, Chan et al. provided in [5] an approximation of the inter-cell interference distribution assuming that the traffic follows a Poisson model. In [6], Haenggi and Ganti provided a good review of interferences in regular and

R. Nasri and A. Jaziri are with Orange Labs, 38/40 avenue General Leclerc, 92794 Issy-les-Moulineaux, France. emails (ridha.nasri@orange.com, aymen.jaziri@orange.com).

random networks and in particular they gave lower and upper bounds of the cumulated interferences in hexagonal lattices assuming that the receiver is placed at the origin [6, p. 19]. In the search of the best approximation that well fits the results of simulations, Karray provided in [3] a comparison between some approximations of ISR in regular hexagonal networks. To make the hexagonal network model tractable, authors in [2] transformed the hexagonal model to the fluid model: interfering sites are continuously and uniformly distributed in the plane. It was shown that the fluid model is a weak approximation of the hexagonal network. Almost all works about interference evaluation in fluid model assume that sites in the network are omni-directional.

Unlike in the hexagonal model, interference analysis is tractable when sites are organized according to a Homogeneous Poisson Point Process (HPPP), i.e., the number of sites in a given area follows a Poisson distribution with constant sites' intensity [6], [8]–[10], [15]. An excellent mathematical theory of interferences in Poisson network of interferers can be found in [6], [8], [9]. Despite its tractability, HPPP model is very irregular and can not always fit with the geometry of real wireless networks because, in reality, sites exhibit repulsive behavior while limiting the dispersion of each site position within a small area having fixed radius. This is related to the fact that radio engineer, while designing sites, places the site at its theoretical hexagonal position in a planning tool, makes coverage prediction and then looks for real candidate sites within a fixed radius from its theoretical position. The deviation of the real network structure from the hexagonal model is therefore linked to the availability of candidate sites (within the search radius) that fulfill certain constraints imposed for example by engineering rules, terrain imperfections and government charters.

To obtain performances of regular networks, while profiting from the tractability of the HPPP model, Haenggi numerically showed in [1] and in [16] with Guo that the SIR and SINR distribution for hexagonal (or even perturbed hexagonal) model with uniform user location can be approximated by shifting the SIR and SINR distribution of the HPPP model. In the same direction, Deng et al. introduced in [17] the Ginibre Point Process that respects a minimum distance between sites and thus ensures more the repulsion of sites, feature naturally identified in real wireless networks.

### C. Contributions

In this work, we treat fundamentally the problem of interference calculation and give explicit and exact formulas of the downlink ISR in hexagonal wireless networks with omni-directional sites. We prove that the ISR metric, as a function of the mobile location $m = re^{i\theta}$, admits an absolutely convergent Fourier series on $\theta$ and an analytic expansion on $r$. The average of the ISR over the angle $\theta$ is rigorously derived without any approximation and involves the use of the well-known Gauss hypergeometric functions [18] and Hurwitz Riemann Zeta functions [19]. Different identities and accurate approximations of ISR are also given and compared with other approximations in literature in [2], [3]. The explicit ISR expression in omni-directional networks is used to derive an accurate formula of ISR in tri-sectorized networks. It is also extended for any frequency reuse pattern. All the provided expressions of the ISR are valid for a propagation parameter $b > 1$.

The second contribution of this work is the explicit derivation of the network performances in terms of SINR distribution in hexagonal networks with an arbitrary user locations' distribution reflecting the spatial traffic density [20]. The finding of SINR distribution entails of course the inversion of the SINR, taken as a function of location $m$. To show the utility of the explicit formula of the SINR distribution, we numerically investigate the SINR distribution for two scenarios of user locations: uniform and heavy-tailed distributions (e.g., Lognormal user location distribution).

### D. Paper organization

The paper is organized as follows. In Section II, system models, notations and preliminary definitions, including the location of omni and tri-sectorized sites, are given. In Section III, we derive the explicit formulas of ISR for omni-directional network and we generalize it for tri-sectorized one. This includes also the presentation of some new approximations and the sketch of their validities and comparison with other approximations in literature. Considering a general frequency pattern and accounting for shadowing effect in the ISR calculation are also provided at the end of section III. In Section IV, analytical expressions of the SINR distributions are given for any user location distribution and numerically presented for uniform and non-uniform user location models. Conclusions summing up important findings are drawn in Section V.

## II. SYSTEM MODELS, NOTATIONS AND DEFINITIONS

### A. Location model of omni-directional sites

We consider a hexagonal wireless network composed of an infinite number of radio sites placed on a regular hexagonal grid as shown in Fig. 1 left part. In this latter, each site is located at the center of the hexagon and composed of only one sector (named also cell). Antenna in each site is assumed to have an omni-directional radiation pattern (uniform antenna gain in all directions) and covers a geographical area, named Voronoi cell. A network with such omni-directional sites is called regular omni-directional network. As largely considered, we assume also that all cells transmit with the same power level $P$.

We identify $\mathbb{R}^2$ with the complex plane $\mathbb{C}$ and we divide it into equally 6 regions $\Omega_l$ ($0 \leq l \leq 5$), defined by

$$\Omega_l = \{m \in \mathbb{C} \mid \frac{l\pi}{3} \leq arg(m) < \frac{(l+1)\pi}{3}\} \quad (1)$$

Interfering sites are arranged into rings of increasing radii. Rings of sites, surrounding the central site located at the origin of $\mathbb{R}^2$, are hexagons. We denote by $k$, $k \geq 1$, the index of each ring in the network. The number of sites in the ring $k$ and belonging to region $\Omega_l$ is $k$. We give so the notation $S_{l,k,j}$ to the site indexed by $j$ in ring $k$ and located in region $\Omega_l$. The site located at the origin of $\mathbb{R}^2$ is denoted formally by $S_{0,0,0}$ but, to simplify notations, we omit the lower index in $S_{0,0,0}$ and use simply $S$. We shall use the same label $S_{l,k,j}$ to refer to the geographical position of the site.

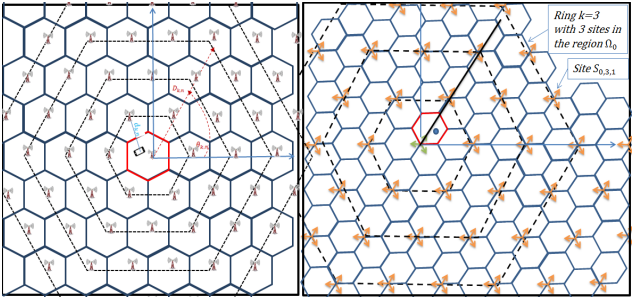

Fig. 1: Hexagonal network model with omni-directional sites (Left side) and tri-sectorized sites (Right side).

*Lemma 1:* In a regular hexagonal network, the site position $S_{l,k,j}$, for $k \geq 1$, is given by

$$S_{l,k,j} = D_{k,j} e^{i(\theta_{k,j} + \frac{l\pi}{3})} \quad (2)$$

where $D_{k,j} = \delta\sqrt{k^2 + j^2 - jk}$, $\theta_{k,j} = atan(\frac{j\sqrt{3}}{2k-j})$, $\delta$ is the inter-site distance and $i = \sqrt{-1}$ is the imaginary unit of the complex plane $\mathbb{C}$.

The proof of the Lemma is simple using the geometry of the network and it is not provided here.

A mobile location in the plane is denoted by the random variable $m$. Likewise, label $m$ is used to denote the geographical location of the mobile. In the complex plane, $m = re^{i\theta}$; where $r = |m|$ is the absolute value of $m$ representing its distance to the central site and $\theta$ is the angle coordinate of the mobile relative to the horizontal axis.

For hexagonal omni-directional network, we assume that location $m$ is connected to central cell $S$ covering a disk of radius $R < \delta$ so that location variable $m \leq R$. The choice of $R$ is problematic because the disk circumscribing the hexagon over-estimates interferences whereas the disk inscribed in the hexagon under-estimates interferences and generates coverage holes. For all the theoretical study, we take an arbitrary $R$, such that $R < \delta$, but for the numerical results we use $R = \delta\sqrt{\frac{\sqrt{3}}{2\pi}}$ which is the radius of the disk having the same area of the hexagon representing the cell. Actually, the real use of $R$ appears only in section IV where we provide the distribution of SINR assuming that $m$ is a random variable having $dt(m)$ as its probability measure.

### B. Location model of tri-sectorized sites

Unlike for omni-directional sites, tri-sectorized sites are located at the corner of the hexagon and composed of 3 sectors covering each one a cell. Sites are arranged to form always a hexagonal grid (see Fig. 1). This model is called regular tri-sectorized network. Since each site is divided into three sectors, we shall use $S_{l,k,j,c}$ to identify the sector $c$, $1 \leq c \leq 3$, of the site $j$ in the ring $k$ and located in region $\Omega_l$. The sector $c$ in the central site is simply denoted $S_c$. The azimuth of the antenna, in which the radiation is at its maximum, is taken relative to the geographical North. Let $\phi_{l,k,j} = arg(m - S_{l,k,j})$ be the angle between site $S_{l,k,j}$ and mobile $m$, then the angle $\phi_{l,k,j,c}$ between the azimuth of sector $c$ in the site $S_{l,k,j}$ and the mobile $m$ is given by

$$\phi_{l,k,j,c} = \frac{\pi}{3}(2c - 3) + \phi_{l,k,j} \quad (3)$$

We denote by $G(.)$ the antenna mask[1] of the sector (relative to its azimuth) and we define the antenna mask of a site, denoted $G_s$, by the sum of all antenna masks of sectors belonging to the same site

$$G_s(\phi_{l,k,j}) = \sum_{c=1}^{3} G(\phi_{l,k,j,c}) \quad (4)$$

### C. Propagation model

In this study, we use simplified Okumura-Hata propagation model [21], i.e., the path-loss between site $S_{l,k,j}$ and mobile user $m$ is expressed as

$$L_{l,k,j}(m) = a\,|m - S_{l,k,j}|^{2b} \quad (5)$$

where $b > 1$ is the amplitude loss exponent, (i.e., $2b$ is the pathloss exponent), $a$ is a constant dependent on the type of the environment (indoor, outdoor, rural, urban,...). We omit also the lower index in $L_{0,0,0}$ and simply use the notation $L$. With this pathloss model and considering the antenna radiation pattern $G$, the received power from sector $S_{l,k,j,c}$ at location $m$ is $\frac{PG(\phi_{l,k,j,c})}{L_{l,k,j}(m)}$, where $\phi_{l,k,j,c}$ is given as in (3).

Since we perform system level analysis, fast fading effect is taken from link level results. This latter feeds the former with link level curve that relates SINR to the throughput or to the packet/bit error rate (see for example [12], [13]). When it is constructed, link level curve accounts for fast fading considering different types of propagation environment and user equipment speeds and categories. Also, the use of fading effect in the link performances finds its motivation from the link adaptation, performed in real networks, in which fast fading is naturally taken in the relation between throughput and SINR. In fact, mobile user sends periodically the channel feedback report in terms of CQI (Channel Quality Indicator reflecting the SINR and including the fading effect) to the enodeB (taking LTE technology for example), and when it is scheduled, the latter chooses the Transport Block size and the Modulation and channel Coding Scheme (MCS) according to the received CQI. So, the throughput is given knowing the CQI (or SINR) and the instantaneous imperfection of the channel due, inter alia, to fast fading.

Moreover, it is known that, for perfect channel (additive White Gaussian Noise Channel), the relation between throughput $th$ (normalized with the bandwidth) and SINR $\xi$ is provided by the famous Shannon's formula $th = \log_2(1 + \xi)$. To account for the imperfection of the channel due to fast fading, one can modify Shannon's formula to have $th = K_1 \log_2(1 + K_2\xi)$, with some constants $K_1$ and $K_2$ calibrated from practical systems as stated in [22].

Shadowing effect is not extensively investigated in this work but we give in section III-D the way of how it can be considered in the ISR calculation. Besides, since the analysis is carried out for arbitrary user location distribution, random variability of the signal, due to shadowing and fast fading, can be aggregated in the user location variable by changing the distribution of user locations accordingly.

---

[1] The antenna mask, $G(\theta) \leq 1$, should not be confused with the antenna gain. In logarithmic (dB) scale, antenna gain is the sum of the maximum antenna gain and the mask.

*D. Definition of ISR*

**Definition** 1: Let $m = re^{i\theta}$ be a location in the plane $\mathbb{C}$ and let $S_{l,k,j}$, as defined in Lemma 1, be the location of a given site in a regular hexagonal network. We denote by $f_{l,k,j}(m)$ the individual ISR received from site $S_{l,k,j}$ in location $m$. It is calculated relatively to the first sector of central site $S$ by

$$f_{l,k,j}(m) = \frac{L(m)}{L_{l,k,j}(m)} \frac{G_s(\phi_{l,k,j})}{G(\theta - \frac{\pi}{3})} \quad (6)$$

Note that for omni-directional site, there is only one sector and hence $G_s = G = 1$.

With the homogeneity hypothesis that all sectors transmit with the same power, $f_{l,k,j}(m)$ is the normalized received power from site $S_{l,k,j}$, against the useful received power at location $m$ served by the first sector of central site $S$.

**Definition** 2: The ISR, denoted by $f(m,b)$, is a function of location $m$ and propagation parameter $b$. It is defined as the sum of all individual ISRs

$$f(m,b) = \sum_{\substack{0 \leq l \leq 5, k \geq 1, \\ 0 \leq j \leq k-1}} f_{l,k,j}(m) \quad (7)$$

## III. MATHEMATICAL ANALYSIS AND EXPLICIT DERIVATION OF THE ISR

In this section, we use mathematical tools to derive the Fourier and Maclaurin series expansion of $f(m,b)$. We first analyze the ISR for omni-directional and tri-sectorized network and we point out, at the end of the section, the possibility to explicitly account for higher frequency reuse pattern and shadowing effect in the ISR calculation.

*A. ISR formulas in hexagonal omni-directional networks*

*1) Explicit and exact formulas of ISR:*

Using the definition of the individual ISR in (6) and substituting $L_{l,k,j}$ by its form in (5), we shall establish the following mathematical results.

***Proposition 1:*** For a given location $m = re^{i\theta}$ connected to the central cell $S$, the individual ISR, defined in (6), is given for omni-directional sites by

$$f_{l,k,j}(m) = \sum_{n=-\infty}^{+\infty} \left(\frac{r}{D_{k,j}}\right)^{2b+|n|} \frac{\Gamma(b+|n|)e^{in(\theta-\theta_{k,j}-l\frac{\pi}{3})}}{\Gamma(b)\Gamma(1+|n|)} \times \\ {}_2F_1\left(b, b+|n|, 1+|n|, \left(\frac{r}{D_{k,j}}\right)^2\right) \quad (8)$$

where ${}_2F_1(.,.,.,.)$ and $\Gamma(.)$ are respectively the Gauss hypergeometric and Euler Gamma functions [18, p. 561].

**Proof:** The proof of proposition 1 is in Appendix A. ∎

***Theorem 1:*** Let $f$, defined as in (7), be the ISR function of a location $m = re^{i\theta}$ and a propagation parameter $b$ in a regular hexagonal network with infinite number of omni-directional sites. Let $\delta$ be the inter-site distance. Take $x = \frac{r}{\delta}$, then for $b > 1$ and $x < 1$, $f$ admits an absolutely convergent Fourier series expansion on $\theta$ and an analytic expansion on $x$

$$f(m,b) = H_0(x,b) + 2\sum_{n=1}^{+\infty} H_n(x,b)\cos(6n\theta) \quad (9)$$

where

$$H_0(x,b) = \frac{6x^{2b}}{\Gamma(b)^2} \sum_{h=0}^{+\infty} \frac{\Gamma(b+h)^2}{\Gamma(h+1)^2} \omega(b+h) x^{2h} \quad (10)$$

$$H_n(x,b) \approx \frac{6\Gamma(b+6n)}{\Gamma(b)\Gamma(1+6n)} \frac{x^{2b+6n}}{(1-x^2)^b} \quad (11)$$

and

$$\omega(b) = 3^{-b}\zeta(b)\left(\zeta(b,\frac{1}{3}) - \zeta(b,\frac{2}{3})\right) \quad (12)$$

with $\zeta(.)$ and $\zeta(.,.)$ are respectively the Riemann Zeta and Hurwitz Riemann Zeta functions [19, p. 1036].

To prove theorem 1, we first establish the following lemma, proved in Appendix B. This Lemma explicitly provides the normalized interferences received at the origin of $\mathbb{C}$ in terms of known functions.

***Lemma 2:*** Let $z$ be any complex number such that $\Re(z) > 1$, where $\Re(z)$ is the real part of the complex number $z$, then

$$\omega(z) = \sum_{k=1}^{+\infty}\sum_{j=0}^{k-1} \frac{1}{(k^2+j^2-jk)^z} \\ = 3^{-z}\zeta(z)\left(\zeta(z,\frac{1}{3}) - \zeta(z,\frac{2}{3})\right) \quad (13)$$

where $\zeta(.)$ and $\zeta(.,.)$ are respectively the Riemann Zeta and Hurwitz Riemann Zeta functions.

*Proof of theorem 1:*

Using the definition of $f(m,b)$ in (7) and the explicit form of $f_{l,k,j}(m)$ in (8), we obtain

$$f(m,b) = \sum_{\substack{0 \leq l \leq 5, k \geq 1, \\ 0 \leq j \leq k-1}} \sum_{n=-\infty}^{+\infty} \left(\frac{r}{D_{k,j}}\right)^{2b+|n|} e^{in(\theta-\theta_{k,j}-l\frac{\pi}{3})} \times \\ \frac{\Gamma(b+|n|)}{\Gamma(b)\Gamma(1+|n|)} \,{}_2F_1\left(b, b+|n|, 1+|n|, \left(\frac{r}{D_{k,j}}\right)^2\right) \quad (14)$$

The sum over $l$ is simple to evaluate because it vanishes for every integer $n$ not multiple of 6. In addition, the infinite hexagonal network is symmetric with respect to the real axis, then $f(m,b)$ stays unchanged when we substitute the location $m$ by its complex conjugate. It follows that $f$ is an even function on $\theta$ and can be written in a closed form

$$f(m,b) = H_0(x,b) + 2\sum_{n=1}^{+\infty} H_n(x,b)\cos(6n\theta) \quad (15)$$

where

$$H_n(x,b) = \frac{6\Gamma(b+6n)}{\Gamma(b)\Gamma(1+6n)} \sum_{k=1}^{+\infty} \sum_{j=0}^{k-1} \left(\frac{x}{D_{k,j}}\right)^{2b+6n} cos(6n\theta_{k,j})$$
$$_2F_1\left(b, b+6n, 1+6n, \left(\frac{x}{D_{k,j}}\right)^2\right) \quad (16)$$

To prove that the Fourier series expansion of $f$ is absolutely convergent, we shall explicitly evaluate $H_n$ and thus show (10) and (11).

We first substitute $D_{k,j}$ with $\delta\sqrt{k^2+j^2-jk}$ and $r$ with $x\delta$. Expanding then Gauss hypergeometric function $_2F_1$ in (16) and using lemma 2 yield

$$H_0(x,b) = \frac{6x^{2b}}{\Gamma(b)^2} \sum_{h=0}^{+\infty} \frac{\Gamma(b+h)^2}{\Gamma(h+1)^2}\omega(b+h)x^{2h} \quad (17)$$

and for $n \geq 1$

$$H_n(x,b) = \frac{6x^{2b+6n}}{\Gamma(b)^2} \sum_{h=0}^{+\infty} \frac{\Gamma(b+h)\Gamma(b+6n+h)}{\Gamma(h+1)\Gamma(1+6n+h)} x^{2h} \times$$
$$\sum_{k=1}^{+\infty} \sum_{j=0}^{k-1} \frac{cos(6n\theta_{k,j})}{(k^2+j^2-jk)^{b+h+3n}} \quad (18)$$

Now when $6n$ is sufficiently high, we have the following equivalences [18, p. 262]:

$$\frac{\Gamma(b+6n+h)}{\Gamma(1+6n+h)} \sim \frac{\Gamma(b+6n)}{\Gamma(1+6n)} \sim (6n)^{b-1} \quad (19)$$

and

$$\sum_{k=1}^{+\infty} \sum_{j=0}^{k-1} \frac{cos(6n\theta_{k,j})}{(k^2+j^2-jk)^{b+h+3n}} \sim 1 \quad (20)$$

Replacing the previous relations (19) and (20) in the expression (18) of $H_n(x,b)$ yields

$$H_n(x,b) \sim \frac{6\Gamma(b+6n)}{\Gamma(b)^2\Gamma(1+6n)} x^{2b+6n} \sum_{h=0}^{+\infty} \frac{\Gamma(b+h)}{\Gamma(h+1)} x^{2h} \quad (21)$$

The last sum in (21) converges uniformly in the unit disk to $\frac{\Gamma(b)}{(1-x^2)^b}$. We obtain then

$$H_n(x,b) \sim \frac{6\Gamma(b+6n)}{\Gamma(b)\Gamma(1+6n)} \frac{x^{2b+6n}}{(1-x^2)^b}, \quad as \ 6n \to +\infty \quad (22)$$

It follows that the Fourier series expansion (9) of the ISR $f$ converges like the series

$$H_0(x,b) + \frac{12x^{2b}}{\Gamma(b)(1-x^2)^b} \sum_{n=1}^{+\infty} \frac{\Gamma(b+6n)}{\Gamma(1+6n)} x^{6n} cos(6n\theta) \quad (23)$$

which converges absolutely for $|x| < 1$. In addition, the ISR $f$ can be very well approximated by the series in (23) for every $x < 1$ and $b > 1$. ∎

It is interesting to note from theorem 1 the following remarks:
- The explicit formulas of the ISR $f(m,b)$ in (9) and its average over $\theta$ in (10) are rapidly convergent series on $m$, for $|m| < \delta$, and were unknown before this work. The first elements of the series are sufficient to capture with a high precision the ISR value at location $m$. Therefore, using these exact expressions is better, in terms of computation time, than performing simulations with high number of sites.
- The received interference at location $m$ is $\mathcal{I}(m,b) = f(m,b)\frac{P}{ar^{2b}}$. At the origin of $\mathbb{R}^2$, i.e. $m=0$, it is simply reduced to the quantity $\mathcal{I}(0,b) = \frac{6P\omega(b)}{a\delta^{2b}}$.
- The Mean Interference-to-Signal Ratio (MISR) for the hexagonal model can be simply deduced by averaging $f(m,b)$ over $m$. When $m$ follows a uniform random variable on the disk of radius $R = \kappa\delta$ (where $\kappa$ could be for example $1/2$, $\sqrt{\frac{\sqrt{3}}{2\pi}}$ or $1/\sqrt{3}$), the MISR is

$$\bar{f}(b) = \frac{6}{\Gamma(b)^2} \sum_{h=0}^{\infty} \frac{\Gamma(b+h)^2\omega(b+h)}{(b+h+1)\Gamma(h+1)^2}\kappa^{2b+2h} \quad (24)$$

which is independent from the intersite distance $\delta$.

*Corollary 1:* Let $f$, as defined in (7), be the ISR function of location $m$ and parameter $b$. Let $x$ be as in theorem 1, then we have for $b > 1$ and $x < 1$

$$f(m,b) \approx H_0(x,b) - \frac{12x^{2b}}{(1-x^2)^b} +$$
$$\frac{2x^{2b}}{(1-x^2)^b} \sum_{l=0}^{5} \Re\left[\left(1 - xe^{i(\theta+l\pi/3)}\right)^{-b}\right] \quad (25)$$

The proof of corollary 1 is simple. We have just to replace $H_n(x,b)$ in the expression of $f(m,b)$ by the approximation of (11), valid for $n \geq 1$, and then evaluate the sum in (23).

**Remark:**
- As can be observed from equation (11), $H_n(x,b)$ decays very fast with $n$ and becomes very small for $n \geq 1$. Thus, the coefficients $H_n(x,b)$, for $n \geq 1$, contribute very little on the value of $f(m,b)$. The ISR $f$ is then a very slowly varying function on $\theta$. This observation was pointed out in [3] based on the curve of $f$.
- The ISR $f(m,b)$ can be well approximated only with $H_0(x,b)$ with an error equal to $O(x^{2b+6})$; where $O$ is the Knuth's notation.

*Corollary 2:* For $b > 1$ and $x < 1$, the following approximations hold.

$$f(m,b) \approx H_0(x,b) \quad (26)$$
$$H_0(x,b) = 6x^{2b}\left(\omega(b) + \omega(b+1)b^2x^2 + O(x^4)\right) \quad (27)$$
$$H_0(x,b) \approx 6x^{2b}\left(_2F_1(b,b,1,x^2) + \omega(b) - 1\right) \quad (28)$$
$$H_0(x,b) \approx 6x^{2b}\left(\frac{1+(1-b)^2x^2}{(1-x^2)^{2b-1}} + \omega(b) - 1\right) \quad (29)$$

*Proof:* Approximation (26) arises directly from the previous remark. Approximation (27) results from the Maclaurin series expansions of $H_0$ provided in (10) and we take only the development to order 2 with respect to $x$.

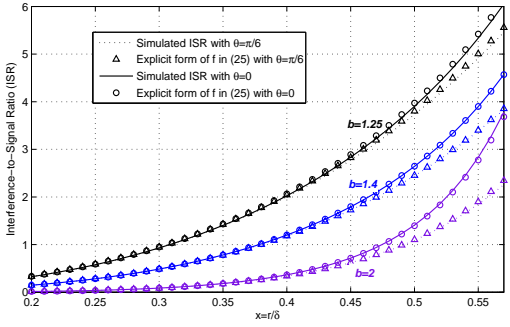

Fig. 2: Explicit formula of $f$ in (25) versus simulated ISR at each location $m$.

To prove (28), let $\omega$ be the function in (13), take the expansion of $H_0$ in (10) and use the approximation $\omega(b+h) \approx 1$, for $h \geq 1$.

To prove (29), we start from (28) and use Euler transformation formula of the Gauss hypergeometric function [18, p. 564]:

$$_2F_1(b,b,1,x^2) = (1-x^2)^{1-2b} {}_2F_1(1-b,1-b,1,x^2) \quad (30)$$

To complete the proof, take only the Maclaurin series development of order 2 in $_2F_1(1-b,1-b,1,x^2)$ because, for $x < 1$ and $b > 1$, the other terms of higher order are negligible. ∎

*2) Validation with numerical examples:*

In Fig. 2, we plot the explicit formula (25) of the ISR and its simulated value in each location $m$ considering different amplitude exponent coefficients $b \in \{1.25, 1.4, 2\}$; and for two cases of angle $\theta \in \{0, \pi/6\}$. For each value of location $m$, the simulated ISR is performed with 1000 rings of cells. We notice that the explicit simple form (25) of $f$ is exactly similar to the one obtained by simulation following its definition in (7), for all $x < 1$ and $b > 1$, but in practice we focus only on the part where the location $m$ is served by the central cell, i.e., $x = \frac{r}{\delta} < \frac{1}{\sqrt{3}}$.

It is worth noting that the ISR undergoes small variations with $\theta$ except at cell edge ($x \approx \frac{1}{\sqrt{3}}$) where its impact becomes significant. Moreover, the influence of $\theta$ on the ISR increases with $b$.

Fig. 3 represents the exact expression of $H_0$ in (10) compared with their development of order 0, namely $6\omega(b)x^{2b}$, and the $2^{nd}$ order given in (27) for $b = 1.1$, $b = 1.3$ and $b = 2$. For $b$ close to 1, both the order 0 and the $2^{nd}$ order match well with the exact form of $H_0$. However, for high value of $b$, the exact value of $H_0$ begins to distance from its first orders' developments and mainly from order 0. This is explained by the fact that when $b$ increases, the coefficient $\frac{\Gamma(b+h)^2}{\Gamma(h+1)^2}$ of the series expansion of $H_0$ also increases. Hence, we need to go for higher order and notably at the cell edge area. To sum up, The $2^{nd}$ order development (27) is a good approximation of $H_0$ in the covered area of the cell for $1 < b \leq 1.5$.

*3) Comparison with other approximations in literature:*

In addition to the explicit calculation (10) of $H_0$, this paper gives a simple approximation of $H_0$ in (29) that would match very well with $H_0$ for every $b > 1$. To show the effectiveness of this approximation, we numerically compare it with other approximations provided in [2] and [3]. The approximation of the ISR (provided for the fluid model) given in [2, Eq.15] is

$$f_1(m) = \frac{2\pi x^{2b}}{\sqrt{3}(b-1)}(1-x)^{2-2b} \quad (31)$$

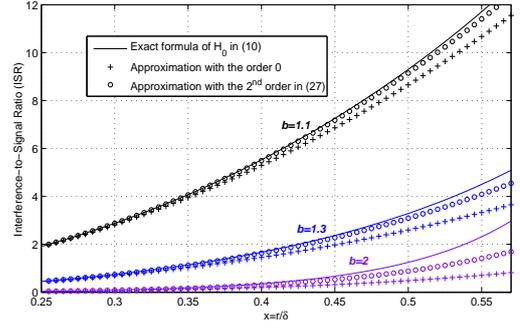

Fig. 3: $H_0$ and its developments of order 0 and $2^{nd}$ order.

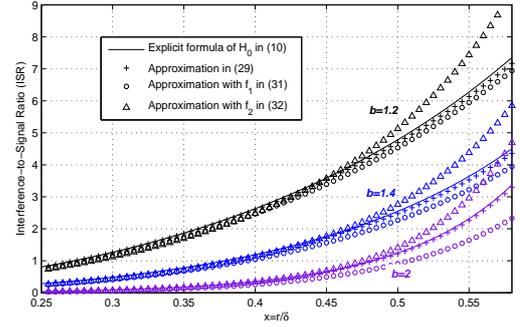

Fig. 4: Comparison of the ISR formula (29) with other approximations for $b \in \{1.2, 1.4, 2\}$.

Whereas, the provided one in [3, Eq.7] is

$$f_2(m) = \frac{\zeta(2b-1)x^{2b}}{(1-x)^{2b}}\left(1 + 4(1-x)^b + \frac{(1-x)^{2b}}{(1+x)^{2b}}\right) \quad (32)$$

In Fig. 4, numerical results show that, for any value of $x \leq \frac{1}{\sqrt{3}}$, the average over $\theta$ of the ISR, $H_0$, is well captured and precisely estimated by the proposed simple approximation (29). The approximation $f_2$, in [3], is valid for low values of $x$ but is limited at the cell edge (with error higher than 20%) for all $b > 1$. The approximation $f_1$, in [2], is good only when $b$ approaches 1. For higher $b$ (e.g., close to 2), $f_1$ moves away from $H_0$ and becomes inefficient to estimate the interferences in hexagonal network. To reduce the gap between $f_1$ and the exact value of $H_0$, Kelif et al. introduced in [2, Eq.16, p. 7] a corrective term obtained by simulation.

### B. ISR expression in hexagonal tri-sectorized networks

Sectorization effect on interferences was largely investigated but mostly with simulations [23], [24]. In regular hexagonal networks with tri-sectorized sites, interferences obviously depend on the antenna radiation patterns of the sectors in both serving and interfering sites. In general, both horizontal and vertical radiation patterns influence interferences, but here we give an approximation of the ISR considering only the horizontal pattern. The vertical one is neglected and we assume that its effect is taken in the constant term of the antenna gain.

As mentioned earlier, antenna pattern (in dB scale) is the sum of a constant gain and a mask depending on the angle between the mobile location and the antenna's boresight.

The antenna mask $G_s$ of a site, as defined in (4), is a periodic function with period $\frac{2\pi}{3}$ if all sectors forming the site have the same antenna pattern. In tri-sectorized network, the relative interference $F$, received in location $m$, is the sum over all interferences received from each site $S_{l,k,j}$ (for $0 \leq l \leq 5$, $k \geq 1$ and $0 \leq j \leq k-1$) and from all collocated sectors of the same site.

$$F(m,b) = -1 + \frac{G_s(\theta)}{G(\theta - \frac{\pi}{3})} + f_s(m,b) \quad (33)$$

where the quantity $-1 + \frac{G_s(\theta)}{G(\theta - \frac{\pi}{3})}$ corresponds to the intra-site ISR generated by sectors belonging to the same site and $f_s(m,b)$ is the inter-site ISR as defined in sub-section II-D. A good approximation of $F(m,b)$ is provided in the following proposition.

***Proposition 2:*** The ISR $F(m,b)$, dependent on $b$ and received at location $m$ connected to the first sector of site $S$ in a regular hexagonal network with infinite number of tri-sectorized sites, has the following approximation:

$$F(m,b) \approx -1 + \frac{G_s(\theta)}{G(\theta - \frac{\pi}{3})} + \frac{\alpha_0 f(m,b)}{G(\theta - \frac{\pi}{3})} - \frac{2\alpha_1 x^{2b}}{G(\theta - \frac{\pi}{3})} \sum_{l=0}^{5} \frac{\Re\left[\left(e^{il\frac{\pi}{3}} - xe^{i\theta}\right)^3\right]}{\left|1 - xe^{i(\theta - l\frac{\pi}{3})}\right|^{2b+3}} \quad (34)$$

where $f$ is recalled the ISR function for omni-directional network, given in theorem 1, and $x = \frac{r}{\delta}$. The coefficients $\alpha_0$ and $\alpha_1$ are given by

$$\alpha_p = \frac{3}{\pi} \int_0^{\frac{\pi}{3}} G_s(\theta) \cos(3p\theta) d\theta, \ p \in \{0,1\} \quad (35)$$

***Proof:*** The proof is provided in Appendix C. ∎

It is important to note that the approach, used to give the closed formula (34) of the ISR in tri-sectorized network, is applicable to any sectorization level such as quadri or hexa-sectorization.

In order to assess the validity of ISR approximation in tri-sectorized network, we use an antenna mask from a real pattern having a beamwidth equal to 65°. The coefficients of the site mask calculated in (35) for the considered antenna are $\alpha_0 = 0.62$ and $\alpha_1 = -0.19$.

In Fig. 5a, we draw both simulated results and the explicit form (34) of the ISR in a hexagonal tri-sectorized network for different values of $b$ and for the extreme values of $\theta$: 0 and $\frac{\pi}{3}$. Simulations is performed using 1000 rings of tri-sectorized sites. For $\theta = 0$, $r \leq \frac{\delta}{3}$ whereas for $\theta = \frac{\pi}{3}$, $r$ varies up to $\frac{2\delta}{3}$. The simple explicit formula (34) approximates very well the ISR for all locations in the cell except with small difference when $m$ approaches $\frac{2\delta}{3}e^{i\frac{\pi}{3}}$ and for high value of $b$. In addition, for $\theta = 0$, the signal coming from the serving sector is almost equal to the signal from the collocated one and consequently ISR is higher than 1 even for locations close to site center. Furthermore, we observe a high gap between ISR in $\theta = 0$ and in $\theta = \frac{\pi}{3}$. Hence, the impact of $\theta$ is no longer negligible

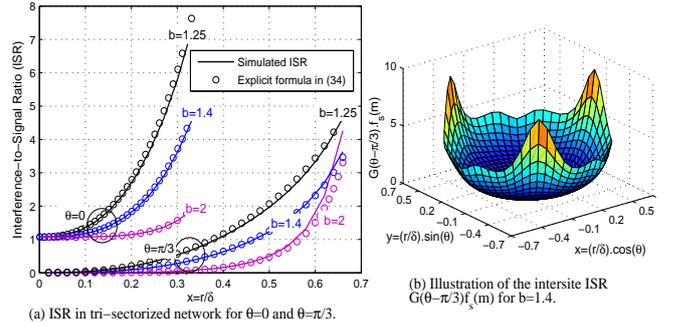

Fig. 5: ISR in tri-sectorized network.

and it plays a crucial role in the calculation of the ISR.

In Fig. 5b, the inter-site ISR $G(\theta - \frac{\pi}{3})f_s(m,b)$, in (59), is represented as a function of $\theta$ and $r$. In contrast to the small impact of $\theta$ in the omni-directional network, we notice further that the inter-site ISR in tri-sectorized network depends on $\theta$ as much as on $r$. In fact, the antenna mask defines the envelope curve of the ISR calculated in omni-directional network. Furthermore, Fig. 5b shows that the inter-site interference is periodic with period $\frac{2\pi}{3}$. This result is easy to be mathematically proved.

### C. ISR expression for higher frequency reuse patterns

Frequency reuse is the core concept of wireless cellular networks and consists in using the same frequency in different geographical areas, spaced far enough apart from each other. To avoid excessive interferences, fixed frequency allocation to each cell is performed in the planning process according to a frequency reuse pattern, denoted here by $v$. According to this pattern, $v$ different frequency bands of the operator are used for each cluster of $v$ adjacent cells.

Co-channel interfering cells are arranged to form again a hexagonal lattice with first ring distant from the cell at the origin by $\Delta = \sqrt{v}\delta$ (see [25, p. 8]), where $\delta$ is the inter-site distance and defined as in section II-A. Let $\hat{f}(m,b,v)$ be the ISR of a hexagonal network having a reuse pattern $v$, it is obtained by changing $\delta$ in the ISR of reuse 1 by $\Delta$. Since the ISR of reuse 1, $f(m,b) = \hat{f}(m,b,1)$, depends on the ratio $m/\delta$, it follows that

$$\hat{f}(m,b,v) = f(m/\sqrt{v}, b) \quad (36)$$

It is clear that $\hat{f}(m,b,v)$ converges faster when $v$ increases. Consequently, for high value of $v$, the first term of the series of $f(m/\sqrt{v}, b)$ is enough to capture the essential value of the ISR. It means that for high value of $v$, $\hat{f}(m,b,v) \approx 6\omega(b)x^{2b}/v^b$, where again $x = |m|/\delta$.

It is known that for high frequency reuse patterns, the operator is obliged to allocate a low bandwidth to each cell resulting in a decrease of cell capacity. In order to maximize the spatial spectrum reuse while minimizing interferences, Fractional Frequency Reuse (FFR) pattern has been proposed in GSM networks for the concentric-cells' deployment (called also Underlay-Overlay cells) and in LTE for the Inter-Cell Interference Coordination (ICIC) scheme. The fundamental idea of FFR is to apply two frequency reuse patterns in the same cell: tight frequency reuse with short reuse distance, restricted to users close to the cell, and regular frequency reuse with long reuse distance, useful for users at cell edge. For FFR, the ISR provided in (36) is still valid but the frequency reuse pattern $v$ depends indeed on the location $m$. A well known example of FFR pattern is $v(m) = \mathbb{1}(|m| \leq r_0) + 3 \times \mathbb{1}(|m| > r_0)$, where $\mathbb{1}(|m| \leq r_0)$ is the indicator function that takes 1 if the condition "$|m| \leq r_0$" is verified and 0 else and $r_0$ is a threshold separating cell edge users from those close to the cell. This means that a frequency reuse pattern $v = 3$ is used at cell edge whereas at cell center $v = 1$.

### D. Inclusion of Log-normal shadowing in the ISR calculation

To take into account Log-normal shadowing effect on the calculation of the ISR, the pathloss $L_{l,k,j}(m)$ between location $m$ and site $S_{l,k,j}$ becomes a Log-normal random variable multiplied by its expression in (5). In each location $m$, $L_{l,k,j}(m)$ and $L_{l',k',j'}(m)$ are also supposed to be independent and identically distributed whenever $(l,k,j) \neq (l',k',j')$. Consequently, the individual ISR, defined in (6), is also a Log-normal random variable since it is the ratio of two Log-normal variables. Let $f^{(sh)}(m,b)$ be the ISR when accounting for the effect of shadowing, its definition in (7) is changed for omnidirectional network by

$$f^{(sh)}(m,b) = \sum_{\substack{0 \leq l \leq 5, k \geq 1, \\ 0 \leq j \leq k-1}} \frac{|m|^{2b}}{|m - S_{l,k,j}|^{2b}} \chi_{l,k,j} \quad (37)$$

where $\{\chi_{l,k,j}\}$ is a sequence of independent and identically distributed Log-normal variables with mean $E[\chi]$ and variance $Var[\chi]$.
The distribution of the Log-normal sum is not exactly known and always constitutes an open problem in probability theory. Nevertheless, it has been recognized that the Log-normal sum can be well approximated by a new Log-normal random variable using Fenton-Wilkinson method [26]. This is achieved by matching the mean and the variance of the new resulted Log-normal variable with those of the Log-normal sum. The choice of Fenton-Wilkinson approach is motivated by its convenience for the present study in contrast with other approaches in [27], [28]. Applying Fenton-Wilkinson method to the Log-normal sum in (37), it follows that $f^{(sh)}(m,b)$ is approximated by a Log-normal random variable with mean

$$E[f^{(sh)}(m,b)] = f(m,b)E[\chi]$$

and variance

$$Var[f^{(sh)}(m,b)] = f(m,2b)Var[\chi]$$

where $f(m,b)$ is the ISR without the inclusion of shadowing and given explicitly in (9) or in (25).
It is important to note that, with the inclusion of shadowing, the Voronoi cell associated to the central cell is no longer the hexagon or the disk having the same area as the hexagon. It becomes the set $\{m \in \mathbb{C} \mid L(m) \leq L_{l,k,j}(m), k \geq 1\}$, instead.

## IV. APPLICATION TO PERFORMANCE ANALYSIS: SINR DISTRIBUTION

The performance analysis of wireless networks definitely entails the evaluation of the SINR distribution. Its knowledge allows to estimate the throughput which is related to the SINR by a continuous, differentiable, and strictly monotonically increasing function. As mentioned in section II-C, this function is called the link level curve of the system.
In this section, we give the explicit formula of the Complementary Cumulative Distribution Function (CCDF) of SINR for a general location distribution. Considering an arbitrary user location distribution is of great interest for capacity dimensioning because it was shown that the spatial user distribution is practically very heterogeneous between cells of the network and even inside cells [20], [29].

### A. Explicit form of SINR distribution

Before the explicit calculation of the SINR CCDF, we give the following definitions.

**Definition** 3: The SINR, experienced at location $m$ connected to central cell $S$ in a regular omni-directional hexagonal network, is the ratio between the useful received power and the total received interferences including thermal noise power

$$SINR = (\eta f(m,b) + y_0 x^{2b})^{-1} \quad (38)$$

where $y_0 = \frac{aP_N}{P}\delta^{2b}$, $P$ is recalled the transmitted power of the cell, including the antenna gain, $a$ and $b$ are the path loss parameters given in (5), $\delta$ is recalled the intersite distance, $x = \frac{r}{\delta}$, $\eta$ is the average load of the interfering cells and $P_N$ is the thermal noise power calculated in the same spectrum bandwidth as for the cell transmit power $P$.
Equation (38) is obtained by firstly considering the classical definition of SINR at location $m$

$$SINR = \frac{P/L(m)}{\sum_{l,k \neq 0,j} \eta_{l,k,j} P/L_{l,k,j}(m) + P_N} \quad (39)$$

where $\eta_{l,k,j}$ is the percentage of occupied resources (scheduling slots or physical resource blocks) in the interfering cell and reflects its load averaged over a period corresponding to the averaging window of the measurement reports of mobiles; and secondly assuming that all interfering cells have the same average load $\eta$.

**Definition** 4: The CCDF of SINR for any scenario of user locations' distribution is given by the following integral

$$\Psi(y) = \mathbb{P}(SINR > y) = \int_S \mathbb{1}(SINR > y) dt(m) \quad (40)$$

where $S$ is the area of the central cell assumed to be a disk of radius $R < \delta$, $\mathbb{1}(.)$ is the indicator function. $t(m)$ is the probability measure of the location variable $m$ and reflects the spatial traffic distribution in cell $S$.
We assume that $t$ is a probability measure on $S$ and consequently

$$\int_S dt(m) = 1$$

**Remark:**
- The CCDF of SINR is widely called coverage probability because it gives the percentage of locations in which the condition "$SINR > y$" is guaranteed.
- When the thermal noise power is neglected with respect to interferences, $y_0 \approx 0$ and the SINR might be similar to SIR.
- When the SINR in each location $m$ is a random variable (e.g., through the shadowing effect), definition (40) of the SINR CCDF is replaced by

$$\Psi(y) = \int_S \mathbb{P}(SINR > y|m) dt(m)$$

In the calculation of the SINR distribution, we consider the expression of the ISR in (26) since we showed earlier that, for an omni-directional network, the angle $\theta$ has a little impact. This latter would be further smoothed when looking for the distribution of any transformation of the ISR function.

Let $g$ be the function defined on $[0, 1)$ by

$$g(x) = \eta H_0(x) + y_0 x^{2b} \tag{41}$$

where again $x = \frac{r}{\delta}$, it is clear that $g(x) = 1/SINR$ and the calculation of SINR CCDF requires the inverse function of $g$ which realizes a continuous bijection from $[0, \rho]$ to $[0, g(\rho)]$ for every real number $\rho$ satisfying $0 \leq \rho < 1$. The inversion of $g$ can be explicitly given by series reversion method. In the following proposition, we provide an approximation of $g^{-1}(y)$.

**Proposition 3:** Let $g$ be defined as in (41). Let $y = g(x)$ for every $x \leq \rho = \frac{1}{\sqrt{3}}$, then the following approximation holds for $x = g^{-1}(y)$

$$g^{-1}(y) \approx \frac{C(y, b)}{\sqrt{\frac{1}{2} + \sqrt{\frac{1}{4} + \beta(b)C(y, b)^2}}} \tag{42}$$

where

$$C(y, b) = \left(\frac{y}{6\eta\omega(b) + y_0}\right)^{\frac{1}{2b}} \text{ and } \beta(b) = \frac{6b\eta\omega(b+1)}{6\eta\omega(b) + y_0}$$

**Proof:** The proof is given in Appendix D. ∎

Considering now an arbitrary traffic distribution, we shall state the explicit form of SINR CCDF involving the inverse function $g^{-1}$ in the following theorem.

**Theorem 2:** Let $g$ be defined as in (41). Recall $\delta$ be the inter-site distance and $R$ be the radius of the central cell $S$. Then the SINR CCDF is explicitly given by

$$\Psi(y) = T(\Lambda(y)), \forall y > 0 \tag{43}$$

with

$$\Lambda(y) = \min\left(\delta \times g^{-1}(\frac{1}{y}), R\right) \tag{44}$$

The function $T(.)$ is the marginal Cumulative Distribution Function of the location random variable $m$ and is obtained from the probability measure $t(m) = t(r, \theta)$ by $dT(r) = \int_0^{2\pi} dt(r, \theta)$. The function $\min(u, v)$ gives the smallest one among the real numbers $u$ and $v$.

**Proof:** Taking the definition of SINR and $\Psi$ in respectively (38) and (40) and with the assumption that the impact of $\theta$ on the ISR is neglected, we have

$$\Psi(y) = \int_0^{2\pi} \int_0^R \mathbb{1}(\frac{1}{g(\frac{r}{\delta})} > y) dt(r, \theta)$$
$$= \int_0^{\min(\delta \times g^{-1}(\frac{1}{y}), R)} \int_0^{2\pi} dt(r, \theta)$$
$$= \int_0^{\Lambda(y)} dT(r)$$

*B. Numerical results for different user locations' distributions*

In order to evaluate and validate the analytical expression of SINR distribution $\Psi$ for different scenarios of user locations' distributions, we consider a uniform and non-uniform locations' distributions in the cell. Note that in practice, the estimation of the spatial traffic density (the measure $t(m)$) in each location can be based on processing probes or on the manipulation of some key performance indicators of the network as in [20]. Furthermore, authors in [29] showed with real data measurements that the spatial traffic density can be approximated by a Log-normal distribution. Therefore, numerical results of this work are provided for both uniform and Log-normal[2] user locations' distribution scenarios.

Besides, for each location scenario, we simulate interferences in a hexagonal omni-directional network using Monte-Carlo method and compare it to the theoretical results. The considered network is composed of 1000 rings of cells operating in the band $2600 Mhz$ of the LTE technology. Only $20Mhz$ of spectrum bandwidth is available and reused everywhere in the network. i.e., frequency reuse pattern equals 1. Each cell of the network transmits with a power equal to $P = 60dBm$[3] and distant from its adjacent cells by $\delta = 1Km$. We set the radius of the cell to $R = \delta\sqrt{\frac{\sqrt{3}}{2\pi}} = 0.525\delta$ and we assume that interfering cells are fully loaded, i.e., $\eta = 1$. The downlink thermal noise[4] is set to $-93dBm$. It is worth noting that parameters in $dB$ scale are transformed to linear when used for the calculation of the simulated metrics ISR and SINR.

For each scenario of simulation, 20000 users are generated according to the considered location distribution in subsections IV-B1 and IV-B2. For each generated user location, interferences, ISR and SINR are calculated.

The comparison between theoretical and simulation results are provided for 2 types of environment, different by the value of the propagation parameter $a$: Outdoor environment with $a = 130dB$ and deep indoor environment with $a = 166dB$. Furthermore, We investigate results for 3 different values of $b$: $1.25$, $1.5$ $and$ $2$.

---

[2]The Log-normal user location distribution should not be confused with the Log-normal shadowing.

[3]$P = 60dBm$ corresponds to $43dBm$ from the transmitter power amplifier and $17dB$ for the antenna gain of the transmitter.

[4]$P_N = E_n + 10log(W) + NF$; where $E_n$ is the energy of the thermal noise equal to $-174dBm/Hz$, $W$ is the spectrum bandwidth equal to $20Mhz$ and $NF$ is the noise figure of the transmitter, set here to $8dB$.

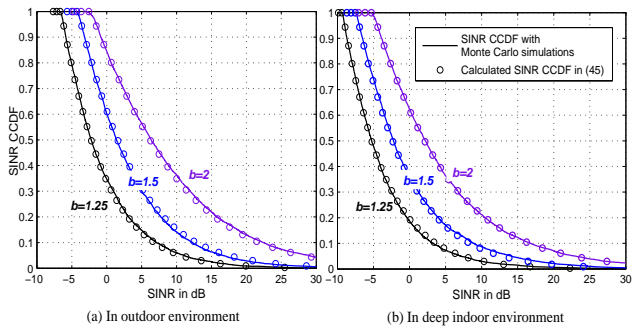

(a) In outdoor environment
(b) In deep indoor environment

Fig. 6: CCDF of the SINR for uniform location distribution scenario with different values of $b$.

*1) SINR distribution for uniform user location variable:*

When the user location distribution is uniform in the disk (representing the cell) of radius $R$, we have $dt(m) = \frac{rdrd\theta}{\pi R^2}$. The distribution $\Psi$ of SINR in (43) becomes

$$\Psi(y) = \frac{\Lambda(y)^2}{R^2} \quad (45)$$

Since the ISR is a function of the polar coordinates of $m$, approximating the central Voronoi cell with a disk of radius $R$ gives a simple and closed form of SINR distribution. This assumption has been widely considered, see for example [30], [31].

In Fig. 6, we present the curve of $\Psi$ in (45) and the SINR CCDF obtained by Monte Carlo simulations for both outdoor environment (Fig.6a) and deep indoor environment (Fig. 6b). The results show that the expression of $\Psi$ calculated in (45) fits with simulations results of the SINR CCDF quite well. Moreover, as expected the impact of $\theta$ on $\Psi$ is smoothed and thus neglected. In addition, it is important to note, from the comparison between Fig. 6a and Fig. 6b, that in the deep indoor environment, the impact of the value $y_0$ on the SINR is clear and tends to decrease it by 2 to $3dB$ depending on $b$. This is explained by the high value of the deep indoor penetration margin that makes the path loss very high and thus the signal arrives very low at indoor locations. Conversely, for outdoor environment, $y_0$ is low and SINR CCDF should be quite similar to SIR CCDF.

*2) SINR distribution for Log-normal user location variable:*
For a Log-normal location distribution, we assume that variable $r$ follows a Log-normal distribution $\sim ln\mathcal{N}(\mu, \sigma)$ with mean $\mu$ and standard deviation $\sigma$. Variable $\theta$ is always uniform $\sim \mathcal{U}(0, 2\pi)$. The probability measure of location $m$ is written by

$$dt(m) = \frac{e^{-\frac{(\ln(r)-\mu)^2}{2\sigma^2}} dr}{\sigma\sqrt{2\pi} Q\left[\frac{1}{\sigma}(\ln(R)-\mu)\right] r} \frac{d\theta}{2\pi}$$

and consequently the expression (43) of $\Psi$ becomes

$$\Psi(y) = \frac{Q\left[\frac{1}{\sigma}(\ln(\Lambda(y))-\mu)\right]}{Q\left[\frac{1}{\sigma}(\ln(R)-\mu)\right]} \quad (46)$$

where $Q$ is the cumulative distribution function of the standard normal distribution and is given by the following integral.

$$Q(z) = \frac{1}{\sqrt{2\pi}} \int_{-\infty}^{z} e^{-\frac{u^2}{2}} du, \ \forall z \in \mathbb{R}$$

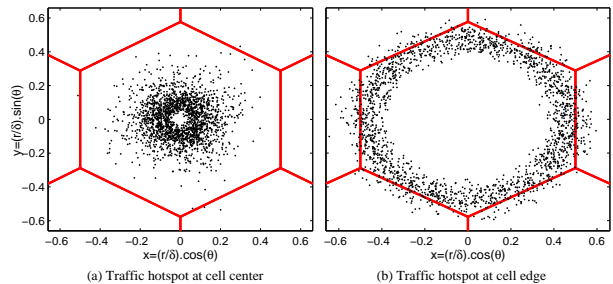

(a) Traffic hotspot at cell center
(b) Traffic hotspot at cell edge

Fig. 7: Snapshots of the two traffic hotspots following Log-normal distribution: The first is close to cell center (a) with $(\mu, \sigma) = (-2, 0.5)$ and the second is at cell edge (b) with $(\mu, \sigma) = (-0.75, 0.1)$.

The parameters $\mu$ and $\sigma$ of the Log-normal distribution allow to deal with different scenarios of traffic hotspots. Besides, $\mu$ and $\sigma$ can be tuned to precisely approximate the distribution of SINR in each cell of a real network.

Here, we show the SINR distribution for two cases of user locations distributions. The first case is a Log-normal hotspot close to cell center with parameters $(\mu, \sigma) = (-2, 0.5)$, whereas the second one is a Log-normal traffic hotspot located at cell edge and having parameters $(\mu, \sigma) = (-0.75, 0.1)$. Both hotspots are depicted in Fig. 7.

In Fig. 8, we show the CCDF of the SINR for the previously cited environments (Fig. 8a for outdoor and Fig. 8b for deep indoor environment) considering a hotspot close to the cell center (Fig. 7a). Simulation Results confirm once again the precision brought by the analytical form of SINR CCDF in (43). We notice indeed that numerical and analytical curves of SINR CCDF are very close. This consolidates also the different observations from the study of uniform location distribution, notably regarding the negligible influence of $\theta$ on the calculation of the SINR distribution and the importance of the penetration margin that deteriorates more or less the SINR in deep indoor environment.

In Fig. 9, we present the SINR CCDF for a Log-normal traffic hotspot located at cell edge (see Fig.7b). Even if $\theta$ influences the ISR value in cell edge (see Fig. 2), the curve of SINR CCDF shows to be still insensitive to $\theta$ for traffic hotspot at cell border because the simulation result considering $\theta$ is almost similar to the analytical expression (43) which does not account for $\theta$ variation. All locations in a hotspot at cell edge suffer from degraded SINR. We note also that the SINR CCDF curve of a hotspot close to cell center is more scattered than that of a hotspot at cell edge. This is related to the value of $\sigma$ which is higher for the first hotspot.

The same previous conclusions regarding the impact of the penetration margin in indoor environment are still valid for hotspot at cell edge. We could say that the SINR CCDF for deep indoor is obtained from that of outdoor environment by a translation of almost $3dB$ to low values.

## V. CONCLUSIONS

This paper has analytically and fundamentally investigated downlink interferences in hexagonal wireless networks. Several results have been established for the ISR function $f$ and the SINR distribution. Given a location $m = re^{i\theta}$, we have proved that the ISR can be written as an absolutely convergent Fourier series on $\theta$ with explicit coefficients. The first coefficient, which is the average of the ISR over $\theta$, has been explicitly evaluated without any approximation. Besides, we have proved for omni-directional network that the ISR is a slowly varying function on $\theta$. Its average, as a function of $r$, is enough to capture the essential information of $f$ when investigating the SINR distribution. In addition to the provided closed formulas of ISR, this paper has also given very simple approximations, rigorously proved and shown to be better than other approximations in literature. Subsequently, the derived expressions have been extended for tri-sectorized hexagonal networks and for the use of high frequency reuse patterns.

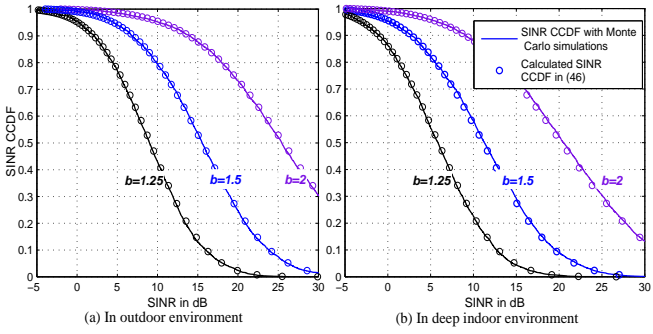

Fig. 8: CCDF of the SINR in a cell having traffic hotspot close to its center:($\mu = -2$ and $\sigma = 0.5$) for different values of $b$.

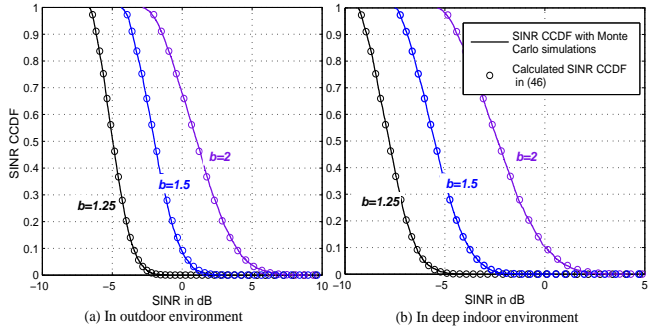

Fig. 9: CCDF of the SINR in a cell having traffic hotspot at cell edge: ($\mu = -0.75$ and $\sigma = 0.1$) for different values of $b$.

Moreover, SINR distribution has been explicitly given for every location distribution and entails, of course, the inversion of the function $g : x \mapsto \eta H_0(x, b) + y_0 x^{2b}$. Numerical results have been carried out for two types of user locations: uniform and Log-normal. The exact formula of SINR distribution approximates quite well the simulated SINR even when user locations form a traffic hotspot at cell edge.

For future works, we will extend the results of this paper to mathematically analyze interferences in perturbed hexagonal network model: site location is perturbed by a random variable. Apart from the fact that any new result for such a network model is scientifically interesting, there are additional motivations for looking on it:

- Perturbed hexagonal network model approximates more than any other model the geometry of the real network [10]. Consequently, performances of perturbed hexagonal model would be closer to the performances of a given real network,
- In addition to the parameters of the location random variable $m$, considered in this paper, the perturbed model introduces a new parameter, named the average perturbation. SINR or throughput distribution for a given real cell can be precisely determined with tuning the average perturbation and the first moments of the user location variable.

## APPENDIX A
## PROOF OF PROPOSITION 1

We first replace $L_{l,k,j}(m)$ in (6) by its expression in (5) and use (2), then we can write down $f_{l,k,j}(m) = \tau(m)\overline{\tau(m)}$; where

$$\tau(m) = \left(\frac{r}{D_{k,j}}\right)^b \left(1 - \frac{r}{D_{k,j}} e^{i(\theta - \theta_{k,j} - l\frac{\pi}{3})}\right)^{-b} \quad (47)$$

and $\overline{\tau(m)}$ is its complex conjugate.

Since $\frac{r}{D_{k,j}} < 1$, $\tau(m)$ is expanded to an absolutely convergent series

$$\tau(m) = \sum_{n=0}^{+\infty} \left(\frac{r}{D_{k,j}}\right)^{n+b} e^{in(\theta - \theta_{k,j} - l\frac{\pi}{3})} \frac{\Gamma(b+n)}{\Gamma(b)\Gamma(1+n)} \quad (48)$$

Arranging the product of the two series $\tau(m)\overline{\tau(m)}$ in order to get a Fourier series expansion on $\theta$ and using the definition of the Gauss hypergeometric function [18, p. 561], we arrive at (8).

## APPENDIX B
## PROOF OF LEMMA 2

Let $z$ be any complex number such that $\Re(z) > 1$, define

$$\omega(z) = \sum_{k=1}^{+\infty} \sum_{j=0}^{k-1} \frac{1}{(k^2 + j^2 - jk)^z} \quad (49)$$

Observe that $\omega(z)\Gamma(z)$ is the Mellin transform of the function $\vartheta$ defined for all complex number $t$, such that $\Re(t) > 0$, by

$$\vartheta(t) = \sum_{k=1}^{+\infty} \sum_{j=0}^{k-1} e^{-t(k^2+j^2-jk)} = \sum_{k=1}^{+\infty} \sum_{j=0}^{+\infty} e^{-t(k^2+j^2+jk)} \quad (50)$$

Now using the Borwein cubic theta function defined by the bilateral sum [32], we have

$$\sum_{k,j \in \mathbb{Z}} e^{-t(k^2+j^2+jk)} = 1 + 2\vartheta(t) + 2\sum_{k=1}^{+\infty}\sum_{j=0}^{+\infty} e^{-t(k^2+j^2-jk)} \quad (51)$$

Dividing the last term in the right hand side of (51) into two parts and manipulating the indexes of the sum give the following identity

$$\sum_{k=1}^{+\infty}\sum_{j=0}^{+\infty} e^{-t(k^2+j^2-jk)} = 2\vartheta(t) \quad (52)$$

It follows, from the last identity and (51), that

$$\sum_{k,j\in\mathbb{Z}} e^{-t(k^2+j^2+jk)} = 1 + 6\vartheta(t) \tag{53}$$

Using the Lambert Series form of Borwein cubic theta function [32], we obtain

$$\vartheta(t) = \sum_{k=0}^{+\infty}\left(\frac{1}{e^{t(3k+1)}-1} - \frac{1}{e^{t(3k+2)}-1}\right) \tag{54}$$

Finally, applying the Mellin transform to both right and left hand sides of the last equation yields

$$\omega(z)\Gamma(z) = 3^{-z}\zeta(z)\left(\zeta(z,\frac{1}{3})-\zeta(z,\frac{2}{3})\right)\Gamma(z) \tag{55}$$

which completes the proof of lemma 2.

## APPENDIX C
## PROOF OF PROPOSITION 2

Let $G_s$ be defined as in (4). Given that $G_s$ is a periodic function on $\theta$ with period $\frac{2\pi}{3}$, we assume that it has a convergent Fourier series expansion

$$G_s(\theta) = \alpha_0 + 2\sum_{p=1}^{+\infty}\alpha_p cos(3p\theta) \tag{56}$$

where

$$\alpha_p = \frac{3}{\pi}\int_0^{\frac{\pi}{3}} G_s(\theta)cos(3p\theta)d\theta,\ \forall p\in\mathbb{Z}$$

Let again $m$ be an arbitrary location in $S$ and recall that $e^{i\phi_{l,k,j}} = (m-S_{l,k,j})/|m-S_{l,k,j}|$. Using the expression of pathloss in (5), we have

$$\frac{L(m)G_s(\phi_{l,k,j})}{L_{l,k,j}(m)} = r^{2b}\sum_{p\in\mathbb{Z}}\alpha_p\frac{(m-S_{l,k,j})^{3p}}{|m-S_{l,k,j}|^{2b+3p}} \tag{57}$$

Thanks to the convergence of the Fourier series (56), the coefficient $\alpha_p$ can be neglected for $p \geq 2$. Consequently, we have

$$\frac{L(m)G_s(\phi_{l,k,j})}{L_{l,k,j}(m)} \approx \alpha_0\frac{L(m)}{L_{l,k,j}(m)} + 2\alpha_1 r^{2b}\frac{\Re\left[(m-S_{l,k,j})^3\right]}{|m-S_{l,k,j}|^{2b+3}} \tag{58}$$

We now sum all individual ISRs arriving at location $m$. For the second term having the coefficient $\alpha_1$, we limit the sum only to the first ring $k=1$ because it behaves like an ISR with path loss exponent equals $2b+3$. To prove the latter, we develop $r^{2b}\Re\left[(m-S_{l,k,j})^3\right]|m-S_{l,k,j}|^{-2b-3}$ as we did in proposition 1. It follows that

$$G(\theta-\frac{\pi}{3})f_s(m,b) \approx \sum_{\substack{0\leq l\leq 5, k\geq 1,\\ 0\leq j\leq k-1}}\frac{L(m)}{L_{l,k,j}(m)}G_s(\phi_{l,k,j})$$

$$\approx \alpha_0 f(m,b) + 2\alpha_1 r^{2b}\sum_{l=0}^{5}\frac{\Re\left[(m-S_{l,1,0})^3\right]}{|m-S_{l,1,0}|^{2b+3}} \tag{59}$$

Replacing $S_{l,1,0}$ by $\delta e^{il\frac{\pi}{3}}$ and $m$ by $re^{i\theta}$, we complete the proof.

## APPENDIX D
## PROOF OF PROPOSITION 3

Let $g$ be the function defined on $[0,1)$ by (41). Using the series expansion of $H_0(x)$ given in (10), it is clear that $g$ admits an analytic expansion on $x = r/\delta$ by

$$g(x) = (6\eta\omega(b)+y_0)x^{2b}\left(1+\sum_{n=1}^{+\infty}\gamma_n x^{2n}\right) \tag{60}$$

where

$$\gamma_n = \frac{6\eta\Gamma(b+n)^2\omega(b+n)}{(6\eta\omega(b)+y_0)\Gamma(b)^2\Gamma(n+1)^2} \tag{61}$$

Let $y = g(x)$, equation (60) can be transformed to

$$\left(\frac{y}{6\eta\omega(b)+y_0}\right)^{\frac{1}{b}} = x^2\left(1+\sum_{n=1}^{+\infty}\gamma_n x^{2n}\right)^{\frac{1}{b}}$$

$$= x^2 + \frac{\gamma_1}{b}x^4 + O(x^6) \tag{62}$$

where $O$ is the Knuth's notation. The second approximation is justified for $x$ sufficiently small. e.g., valid for $x < \frac{1}{\sqrt{3}}$. Omitting the error terms in equation (62) gives a second-order equation, in $x^2$, which admits two solutions:

$$x_{\pm}^2 = \frac{C(y,b)^2}{\frac{1}{2}\pm\sqrt{\frac{1}{4}+\frac{\gamma_1}{b}C(y,b)^2}} \tag{63}$$

where $C(y,b) = \left(\frac{y}{6\eta\omega(b)+y_0}\right)^{\frac{1}{2b}}$.

Since $x = r/\delta$ is a positive real number, only the positive solution $x_+^2$ is valid. Now, replacing $\gamma_1$ by its expression in (61), we obtain the approximation of $x = g^{-1}(y)$ in proposition 3.

## ACKNOWLEDGMENT

The authors thank the editor and the anonymous reviewers for their constructive comments. They also gratefully Acknowledge Prof. Martin Haenggi for his useful suggestions improving the quality of the paper.

## REFERENCES

[1] M. Haenggi, "The mean interference-to-signal ratio and its key role in cellular and amorphous networks," *Wireless Communications Letters, IEEE*, vol. 3, no. 6, pp. 597–600, 2014.
[2] J.-M. Kelif, M. Coupechoux, and P. Godlewski, "A fluid model for performance analysis in cellular networks," *EURASIP Journal on Wireless Communications and Networking*, vol. 2010, p. 1, 2010.
[3] M. K. Karray, "Study of a key factor for performance evaluation of wireless cellular networks: The f-factor," in *2nd IFIP Wireless Days (WD)*. IEEE, 2009, pp. 1–6.
[4] M. Castaneda, M. T. Ivrlač, J. Nossek, I. Viering, A. Klein *et al.*, "On downlink intercell interference in a cellular system," in *IEEE 18th International Symposium on Personal, Indoor and Mobile Radio Communications, PIMRC 2007*. IEEE, 2007, pp. 1–5.
[5] C. C. Chan and S. V. Hanly, "Calculating the outage probability in a CDMA network with spatial poisson traffic," *Vehicular Technology, IEEE Transactions on*, vol. 50, no. 1, pp. 183–204, 2001.
[6] M. Haenggi and R. K. Ganti, *Interference in large wireless networks*. Now Publishers Inc, 2009.
[7] A. J. Viterbi, A. M. Viterbi, and E. Zehavi, "Other-cell interference in cellular power-controlled CDMA," *Communications, IEEE Transactions on*, vol. 42, no. 234, pp. 1501–1504, 1994.

[8] J. G. Andrews, F. Baccelli, and R. K. Ganti, "A tractable approach to coverage and rate in cellular networks," *Communications, IEEE Transactions on*, vol. 59, no. 11, pp. 3122–3134, 2011.

[9] M. Z. Win, P. C. Pinto, L. Shepp *et al.*, "A mathematical theory of network interference and its applications," *Proceedings of the IEEE*, vol. 97, no. 2, pp. 205–230, 2009.

[10] A. Guo and M. Haenggi, "Spatial stochastic models and metrics for the structure of base stations in cellular networks," *Wireless Communications, IEEE Transactions on*, vol. 12, no. 11, pp. 5800–5812, 2013.

[11] D. W. Matolak and A. Thakur, "Outside cell interference dynamics in cellular CDMA," in *Proceedings of the 35th Southeastern Symposium on System Theory*. IEEE, 2003, pp. 148–152.

[12] J. C. Ikuno, M. Wrulich, and M. Rupp, "System level simulation of LTE networks," in *IEEE 71st Vehicular Technology Conference (VTC 2010-Spring)*. IEEE, 2010, pp. 1–5.

[13] F. Brouwer, I. De Bruin, J. C. Silva, N. Souto, F. Cercas, and A. Correia, "Usage of link-level performance indicators for HSDPA network-level simulations in e-UMTS," in *IEEE Eighth International Symposium on Spread Spectrum Techniques and Applications*. IEEE, 2004, pp. 844–848.

[14] A. Jaziri, R. Nasri, and T. Chahed, "System level analysis of heterogeneous networks under imperfect traffic hotspot localization," *To appear in IEEE Transactions on Vehicular Technology*, 2016.

[15] R. Nasri and A. Jaziri, "Tractable approach for hexagonal cellular network model and its comparison to poisson point process," in *Proc. IEEE Globecom 2015, Wireless Communications Symposium*, 2015.

[16] A. Guo and M. Haenggi, "Asymptotic deployment gain: A simple approach to characterize the SINR distribution in general cellular networks," *Communications, IEEE Transactions on*, vol. 63, no. 3, pp. 962–976, 2015.

[17] N. Deng, W. Zhou, and M. Haenggi, "The Ginibre point process as a model for wireless networks with repulsion," *Wireless Communications, IEEE Transactions on*, vol. 14, no. 1, pp. 107–121, 2015.

[18] M. Abramowitz and I. A. Stegun, *Handbook of mathematical functions: with formulas, graphs, and mathematical tables*. Courier Corporation, 1964, no. 55.

[19] A. Jeffrey and D. Zwillinger, *Table of integrals, series, and products*, 7th ed. Elsevier Academic Press publications, 2007.

[20] A. Jaziri, R. Nasri, and T. Chahed, "Traffic hotspot localization in 3G and 4G wireless networks using OMC metrics," in *IEEE International Symposium on Personal, Indoor and Mobile Radio Communications (PIMRC 2014)*. IEEE, 2014, pp. 246–249.

[21] M. Hata, "Empirical formula for propagation loss in land mobile radio services," *Vehicular Technology, IEEE Transactions on*, vol. 29, no. 3, pp. 317–325, 1980.

[22] P. Mogensen, W. Na, I. Z. Kovács, F. Frederiksen, A. Pokhariyal, K. Pedersen, T. Kolding, K. Hugl, M. Kuusela *et al.*, "LTE capacity compared to the shannon bound," in *IEEE 65th Vehicular Technology Conference, VTC2007-Spring*. IEEE, 2007, pp. 1234–1238.

[23] G. K. Chan, "Effects of sectorization on the spectrum efficiency of cellular radio systems," *Vehicular Technology, IEEE Transactions on*, vol. 41, no. 3, pp. 217–225, 1992.

[24] I. Riedel and G. Fettweis, "Increasing throughput and fairness in the downlink of cellular systems with N-fold sectorization," in *IEEE GLOBECOM Workshops (GC Wkshps)*. IEEE, 2011, pp. 167–172.

[25] V. Vij, *Wireless Communication*. Laxmi Publications, Ltd., 2010.

[26] L. F. Fenton, "The sum of log-normal probability distributions in scatter transmission systems," *Communications Systems, IRE Transactions on*, vol. 8, no. 1, pp. 57–67, 1960.

[27] S. C. Schwartz and Y.-S. Yeh, "On the distribution function and moments of power sums with log-normal components," *Bell System Technical Journal*, vol. 61, no. 7, pp. 1441–1462, 1982.

[28] N. B. Mehta, J. Wu, A. F. Molisch, and J. Zhang, "Approximating a sum of random variables with a lognormal," *Wireless Communications, IEEE Transactions on*, vol. 6, no. 7, pp. 2690–2699, 2007.

[29] D. Lee, S. Zhou, X. Zhong, Z. Niu, X. Zhou, and H. Zhang, "Spatial modeling of the traffic density in cellular networks," *Wireless Communications, IEEE*, vol. 21, no. 1, pp. 80–88, 2014.

[30] K. B. Baltzis, *Hexagonal vs circular cell shape: a comparative analysis and evaluation of the two popular modeling approximations*. INTECH Open Access Publisher, 2011.

[31] H. Tabassum, F. Yilmaz, Z. Dawy, and M.-S. Alouini, "A framework for uplink intercell interference modeling with channel-based scheduling," *Wireless Communications, IEEE Transactions on*, vol. 12, no. 1, pp. 206–217, 2013.

[32] J. M. Borwein and P. B. Borwein, "A cubic counterpart of Jacobi's identity and the AGM," *Transactions of the American Mathematical Society*, pp. 691–701, 1991.